\documentstyle[12pt]{article}

\newcommand{\be}{\begin{equation}}
\newcommand{\ee}{\end{equation}}
\newcommand{\bea}{\begin{eqnarray}}
\newcommand{\eea}{\end{eqnarray}}
\newcommand{\nn}{\nonumber \\}

\newcommand{\ba}{\begin{array}}
\newcommand{\ea}{\end{array}}
\newcommand{\vs}[1]{\vspace{#1 mm}}

\def\bbox{{\,\lower0.9pt\vbox{\hrule \hbox{\vrule height 0.2 cm
\hskip 0.2 cm \vrule height 0.2 cm}\hrule}\,}}
\newcommand{\dsl}{\pa \kern-0.5em /}

\newcommand{\pa}{\partial}

\font\mybb=msbm10 at 12pt
\def\bb#1{\hbox{\mybb#1}}

\def\bR {\bb{R}}

\begin{document}

\topmargin 0pt
\oddsidemargin 5mm

\renewcommand{\thefootnote}{\fnsymbol{footnote}}
\begin{titlepage}

\setcounter{page}{0}
\begin{flushright}
FTUV-98/73 IFIC-98/74 DAMTP-1998-136\\
Oct. 27 (Revised Nov. 2), 1998\\
hep-th/9810230
\end{flushright}

\vs{5}
\begin{center}
{\Large SUPERCONFORMAL MECHANICS, BLACK HOLES, AND NON-LINEAR REALIZATIONS}
\vs{10}

{\large
J.A. de Azc{\'a}rraga$^1$, J.M. Izquierdo$^2$, J.C. P{\'e}rez Bueno$^1$ \\
and  P.K. Townsend$^3$\footnote{E-mails: azcarrag@lie1.ific.uv.es,
izquierd@fta.uva.es, pbueno@lie.ific.uv.es,\\
pkt10@damtp.cam.ac.uk.}}\\
\vs{5}
${}^1${\em Departamento de F\'{\i}sica Te{\'o}rica, Universidad de Valencia\\
       and IFIC, Centro Mixto Universidad de Valencia--CSIC\\
      E--46100 Burjassot (Valencia), Spain}\\
${}^2${\em Departamento de F\'{\i}sica Te\'orica, Universidad de
         Valladolid\\
       E--47011, Valladolid, Spain}\\
${}^3$ {\em DAMTP, University of Cambridge\\
Silver Street, Cambridge, CB3 9EW, UK}
\end{center}
\vs{7}
\centerline{{\bf Abstract}}

The $OSp(2|2)$-invariant planar dynamics of a $D=4$ superparticle near the
horizon of a large mass extreme black hole is described by an $N=2$
superconformal mechanics, with the $SO(2)$ charge being the superparticle's
angular momentum. The {\it non-manifest} superconformal invariance of the
superpotential term is shown to lead to a shift in the $SO(2)$ charge by the
value of its coefficient, which we identify as the orbital angular momentum.
The full $SU(1,1|2)$-invariant dynamics is found from an extension to $N=4$ 
superconformal mechanics.  

\vskip 1cm

\end{titlepage}
\newpage
\renewcommand{\thefootnote}{\arabic{footnote}}
\setcounter{footnote}{0}

\section{Introduction and the $OSp(1|2)$ model}

The dynamics of a particle described by the action
\begin{equation} \label{1.1}
 I=\int dt\, \left[ \frac{1}{2} {m\dot x}^2-\frac{g}{2x^2}\right]
\end{equation}
is invariant under the group $SL(2,\bR)$, one of the generators
being the Hamiltonian $H$. The group $SL(2,\bR)$ is the conformal
group in a `spacetime' of one dimension (time) so the action $I$ is that
of a one-dimensional conformal `field' theory, \emph{i.e.} a model of conformal
mechanics. The model was introduced, and its quantum properties
investigated, in \cite{DFF}. Recently, it was shown that it
describes the radial motion of a particle of mass $m$ and
charge $q$ near the horizon of an extreme (\emph{i.e.} $M=|Q|$)
Reissner-Nordstr{\"o}m
(RN) black hole in a limit in which $|q|/m$ tends to unity at the same time
as the black hole mass $M$ tends to infinity, with $M^2(m-|q|)$
remaining finite \cite{claus}. The coupling constant $g$ is then found to be
\begin{equation}\label{ccrel}
g=8M^2(m-|q|) + 4\ell(\ell +1)/m \quad ,
\end{equation}
where $\ell$ is the particle's orbital angular momentum\footnote{When
$\ell\ne0$ the particle's motion is not purely radial, of course, but by
`radial motion' above we mean the equation for the radial position of the
particle.}.

It was also shown in \cite{claus} that the radial motion of a
superparticle in the same background, but with zero angular momentum, is
described, in the same limit, by an $OSp(1|2)$-invariant \emph{superconformal}
mechanics. However, because the fermionic gauge  symmetries of the
superparticle
require $m=|q|$, and because $\ell=0$ is assumed, the coupling constant
$g$ of this model vanishes, and the potential term is therefore absent. This is
a reflection of the exact balance of the gravitational and electric forces on a
\emph{static} superparticle in an extreme RN black hole background. It was
further pointed out in \cite{claus} that the full superparticle dynamics must
be invariant under the larger $SU(1,1|2)$ superconformal group because this is
the isometry group of the $adS_2\times S^2$ near-horizon supergeometry. This full
dynamics will of course describe not only the radial motion of the
superparticle but also its motion on the 2-sphere. However, there is nothing to
prevent us from considering only the radial motion, which will be the equation
of motion of an $SU(1,1|2)$--invariant generalisation of (\ref{1.1}).

In addition to considering only the radial equation of motion of the
superparticle we can also consider a restriction on the full dynamics in which
the particle is assumed to move within an equatorial plane, or the further
restriction to purely radial motion (\emph{i.e.} $\ell=0$). These restrictions
correspond to a reduction of the superconformal symmetry to some subgroup
of $SU(1,1|2)$, in fact to the sequence of subgroups
\begin{equation}
SU(1,1|2) \supset SU(1,1|1)\cong OSp(2|2) \supset OSp(1|2)\, .
\end{equation}
In the first restriction, to $SU(1,1|1)$, the $SU(2)$ group of rotations is
reduced to the $U(1)$ group of rotations in the plane. The corresponding
superconformal mechanics is the $SU(1,1|1)$ generalisation of (\ref{1.1})
constructed and analysed in \cite{akulov,Fub}. As the above discussion
suggests, the $U(1)$ charge of this model is directly related to the angular
momentum of a superparticle; the precise relation will be given below. That the
subsequent restriction to $OSp(1|2)$ describes purely radial motion was
justified in detail in \cite{claus}. Of principal interest here are the
$OSp(2|2)$ and $SU(1,1|2)$ models because they allow $\ell\ne0$ and hence
$g\ne0$.

The $OSp(2|2)$ superconformal mechanics was initially presented as
a particular model of $N=2$ supersymmetric quantum mechanics \cite{akulov,Fub}.
Its superspace action is a functional of a single worldline superfield
$x(t,\eta_1,\eta_2)$, where $\eta_i\ (i=1,2)$ are anticommuting partners to the
worldline time coordinate $t$. This action is\footnote{The constant $\ell$ is
related to the constant $f$ of \cite{Fub} by $f=2\ell$.}
\begin{equation}
I=-i\int dt\, d^2\eta \left\lbrace 2m
iD_1 x D_2x + 4 \ell \log x\right\rbrace
\quad,
\label{1.2}
\end{equation}
where
\begin{equation}
D_{\eta_i}\equiv D_i = {\partial\over {\partial \eta_i}} -
{i\over2} \eta_i {\partial\over{\partial t}}
\quad,\quad
i,j=1,2\quad,\quad
\left\lbrace D_{i} , D_{j} \right\rbrace = -i \delta_{ij} \partial_t
\quad,
\label{1.3}
\end{equation}
are the superworldline covariant spinor derivatives.

The action (\ref{1.2}) is manifestly invariant under the two worldline
supersymmetries, with corresponding Noether charges $Q_i$, but is also
invariant under the other two supersymmetries of $OSp(2|2)$, with
Noether charges $S_i$. The full set of Noether charges includes those
corresponding to  dilatations ($D$), proper conformal transformations ($K$),
and the $so(2)$ charge {$B$}. For $\ell=0$ these Noether charges obey the
(anti)commutation relations of the $osp(2|2)\cong su(1,1|1)$ algebra.
The non--zero (anti)commutators are
\begin{equation}
\begin{array}{lll}
[H,D]=iH \quad, & [K,D]=-iK \quad, & [H,K]=2iD \quad,
\\[0.3cm]
\{Q_i,Q_j\}=\delta_{ij} H \quad,
& \{S_i,S_j\}=\delta_{ij} K \quad, &
\displaystyle \{Q_i,S_j\}=\delta_{ij} D + \frac 12
\epsilon_{ij} B \quad,
\\[0.3cm]
\displaystyle [D,Q_i]= -\frac i2 Q_i \quad,
& \displaystyle [D,S_i]= \frac i2 S_i \quad,
\\[0.3cm]
[K,Q_i]= -i S_i \quad, & [H,S_i]= i Q_i \quad,
\\[0.3cm]
[B,Q_i]= -i \epsilon_{ij} Q_j \quad,
& [B,S_i]= -i \epsilon_{ij} S_j \quad, & i,j=1,2 \quad .
\end{array}
\label{1.4}
\end{equation}
When $\ell \neq 0$ one finds the same algebra but $B$ is no longer the
$U(1)$ Noether charge associated to the $U(1)$ invariance of (\ref{1.2}).
This is not due to any change in this Noether charge,
which continues to be the same fermion bilinear as before (given by eqn.
(\ref{eq75}) below).
Let us use $\hat B$ to denote this fermion bilinear.
Then, $B$ in (\ref{1.4}) is given by $B=\hat B +2\ell$, so $B=\hat B$
when $\ell=0$ but not otherwise. The main aim of this paper is to provide a
mathematical explanation for why this shift of the $U(1)$ charge occurs, and a
physical explanation of its significance.

The $Q_i$-supersymmetries are linearly realized by the action (\ref{1.2}).
The $S_i$-supersymmetries are non-linearly realized, the variables
$(D_i\,x)|$ being the corresponding Goldstone fermions (where, as usual,
$|$ is short for $|_{\eta_i =0}$). Thus, the above supersymmetric mechanics is
one in which supersymmetry  is partially broken. In fact, the supersymmetry is
`half--broken', as is to be expected from its superparticle origin, and
$x(t,\eta)$ is the Nambu-Goldstone superfield. Since the terms in the action of
lowest  dimension should be determined entirely by (super)symmetry we may use
the method of non-linear realizations of spacetime (super)symmetries
\cite{nonlinear} to construct them. This was done in \cite{IKL} for a class of
$SU(1,1|n)$-invariant $N=2n$ superconformal mechanics models that include the
bosonic model and the $N=2,4$ models of interest here. As a simple illustration
of the method we shall now show how the $OSp(1|2)$-invariant $N=1$
superconformal mechanics of
\cite{claus} can be found in this way.

The superalgebra $osp(1|2)$ is spanned by ($H, K, D, Q, S)$. We choose as an
`unbroken' subalgebra that spanned by $H$ and $Q$, to which we
associate the independent variables $\zeta^M=(t,\eta)$. These parametrise a
real (1,1)--dimensional superworldline. The $OSp(1|2)$ group element on the
superworldline is written as
\begin{equation}
g(t,\eta)= e^{-itH} e^{i\eta Q} e^{i\lambda (t,\eta) S} e^{i z(t,\eta) D}
e^{i\omega (t,\eta) K}
\quad,
\label{2.1}
\end{equation}
where the dependent variables $z$ and $\omega$ associated with the `broken'
generators $D$ and $K$, are commuting worldline superfields, and
$\lambda$ is an anticommuting worldline superfield associated with the `broken'
supercharge $S$. It will prove convenient to introduce the (non-exact)
differential
\begin{equation}
d\tau\equiv dt -\frac{i}{2}\eta d\eta \quad,
\label{2.4a}
\end{equation}
because we then have
\begin{equation}
d= d\tau \partial_t + d\eta D_\eta\quad,
\end{equation}
where
\begin{equation}
{D}_\eta
\equiv \frac\partial{\partial\eta} - \frac{i}{2}\eta \frac\partial {\partial t}
\equiv \partial_\eta - \frac{i}{2}\eta \partial_t \quad ,
\label{2.3a}
\end{equation}
is the superworldline covariant spinor derivative satisfying $2D_\eta^2 =
-i\partial_t$. A calculation now yields\footnote{We assume for the purposes of
this calculation, and those to follow, that $\eta$ and $\lambda$ anticommute
with $Q$ and $S$. The opposite assumption, that they commute, leads to a change
of sign of all fermion bilinears. Since this sign is not fixed by physical
considerations we are free to make either choice for present purposes. We leave
the reader to decide whether one or the other choice is required for
mathematical
consistency.}
\begin{equation}
\begin{array}{r@{}l}
i g^{-1} dg = &
\displaystyle
d\tau e^{-z} H - (d\tau \lambda + d\eta) e^{-z/2} Q
- \bigl[ d\tau(2\omega e^{-z} + \dot z) + d\eta ({D}_\eta z -i\lambda) \bigr] D
\\[0.3cm]
- &
\displaystyle
 \bigl[ d\tau(\dot\omega - \omega\dot z - e^{-z} \omega^2 +
\frac{i}{2}\lambda\dot\lambda e^{z}) + d\eta ({D}_\eta\omega - \omega {D}_\eta
z
+i\omega\lambda
-\frac{i}{2} e^{z} \lambda {D}_\eta\lambda) \bigr] K
\\[0.3cm]
- &
\displaystyle
\bigl[d\tau (\dot\lambda e^{z/2} - \omega e^{-z/2}\lambda) +
d\eta ({D}_\eta e^{z/2} - \omega e^{-z/2})\bigr] S
\quad.
\end{array}
\label{2.2a}
\end{equation}
We can rewrite this in the form\footnote{We shall use 
the caligraphic ${\cal D}$ to denote the \emph{group}
covariant derivatives ${\cal D}_A,\ {\cal D}_M$ used in this paper,
while $D_\eta$ or $D_{\eta_i}\equiv D_i$ will refer to the
\emph{superworldline} derivatives.}
\begin{equation}
i g^{-1} d g = d\xi^M E_M{} ^{~A} [ H_A - ({\cal D}_A z ) D -
({\cal D}_A \omega ) K - ({\cal D}_A \lambda ) S ]
\quad,
\label{2.5}
\end{equation}
where $d\xi^M \equiv (d\tau,d\eta)$ and $H_A =(H_0,H_1)\equiv (H,Q)$. We find
that
\begin{equation}
E_M{}^{A} = \pmatrix { e^{-z} & - \lambda e^{-z/2} \cr
                       0 & -e^{-z/2}}
\quad,\quad
E_A{}^{M} = \pmatrix { e^{z} & - \lambda e^{z} \cr
                       0 & -e^{z/2}}
\quad,
\label{2.6}
\end{equation}
and, for example,
\begin{equation}
{\cal D}_A \omega \equiv
\pmatrix { {\cal D}_0 \omega \cr {\cal D}_1 \omega }
=
\pmatrix {
\displaystyle
e^{z} (\dot\omega -\omega \dot z - e^{-z}\omega^2 +\frac{i}{2}\lambda\dot
\lambda e^{z} -\lambda {D}_\eta\omega+\lambda\omega {D}_\eta z)
\cr
\displaystyle
-e^{z/2} ({D}_\eta\omega-\omega {D}_\eta z +i\omega\lambda -
\frac{i}{2} e^{z} \lambda {D}_\eta\lambda )
}
\quad.
\label{2.7}
\end{equation}

Manifestly $OSp(1|2)$--invariant superspace actions have the form
\begin{equation}
I =\int dt\, d\eta\, (\rm{sdet}\, E)
{\cal L}({\cal D}_A z, {\cal D}_A \omega,{\cal D}_A \lambda)
\quad,
\label{2.8}
\end{equation}
where ${\cal L}$ is an anticommuting worldline scalar. The invariance is 
manifest in the sense that ${\rm sdet} E\, {\cal L}$ transforms as a scalar
density\footnote{Let the infinitesimal transformation of the 
coordinates $\zeta^M=(t,\eta)$ be $\delta\zeta^M
=(\delta t, \delta\eta)$. Then, a scalar density $L$ is one for which $\delta
L= (\delta\zeta^M L){\buildrel \leftarrow \over \partial}_M$.}.
The superspace structure group is chosen so as to leave invariant
the relation $\displaystyle 2D_\eta^2=-i\partial_t$. It follows that the
fermion and boson components of the covariant derivatives are independent tensors
(in fact, scalars in this case). The lowest dimension Lagrangian is therefore
proportional to the ${\cal D}_1 \omega$ component.
All the other choices lead to higher-derivative component
actions\footnote{Assuming that ${\cal D}_Az=0$ is imposed as a constraint to
eliminate $\omega$ and $\lambda$ as independent superfields, because $\omega$
would otherwise be an independent field with wrong-sign kinetic terms.}. Since
\begin{equation}
\rm{sdet}\, E = - e^{-z/2}
\label{2.9}
\end{equation}
we have, discarding a total derivative (the ${D}_\eta\omega$ term),
\begin{equation}
I = -i m\int dt\, d\eta \left( \omega {D}_\eta z - i\omega\lambda +
\frac{i}{2} e^{z} \lambda {D}_\eta\lambda \right)
\quad.
\label{2.10}
\end{equation}
The $\omega$ and $\lambda$ equations yield
\begin{equation}
\omega = - \frac 12 e^{z} \dot z \quad,\quad
\lambda = - i {D}_\eta z
\quad,
\label{2.11}
\end{equation}
which are equivalent to the manifestly $OSp(1|2)$--invariant constraints
${\cal D}_A z=0$, which could have been imposed \emph{ab initio}.
Either way, the action then reduces to
\begin{equation}
I(z) = \frac{i}{4}m\int dt\, d\eta\,  e^{z} \dot z {D}_\eta z\quad.
\label{2.12}
\end{equation}
The equation of motion is equivalent, when combined with the constraint
${\cal D}_A z=0$, to the manifestly $OSp(1|2)$-invariant equation 
${\cal D}_A \omega=0$ (and these imply ${\cal D}_A \lambda=0$).
Setting 
\be
z=\log x^2\, ,
\ee
performing the superspace integral, and then setting the fermions to zero
we recover the Lagrangian (\ref{1.1}) with $g=0$. By retaining the fermions
we recover the $N=1$ superconformal mechanics of \cite{claus}.
 
Note that it is not possible to construct an $OSp(1|2)$--extension of
the $g/x^2$ potential. This might be possible if we  were to suppose that
all supersymmetries are non--linearly realized, but the resulting action would
involve variables other than the components of the superfield $z(t,\eta)$,
and it would not be  expressible in superfield form. To find a suitable
supersymmetric generalization of the potential term we must consider the
further extension to $N=2$ or $N=4$. Similar techniques to those just
described were used in \cite{IKL} to obtain the field equations of the $N=2$
superconformal mechanics model of 
\cite{akulov,Fub} in manifestly $OSp(2|2)\cong
SU(1,1|1)$ invariant form, as a special case of a construction valid for
$SU(1,1|n)$), but no attempt was made to demonstrate {\it manifest} invariance
of the superspace {\it action}. There is a good reason for this: as we shall
see here, the superpotential term of the $SU(1,1|1)$ model {\it cannot be
expressed in a manifestly invariant form}. This possibility arises because
manifest invariance is only a sufficient condition for invariance, not a
necessary one. 

The existence of actions which  are invariant but not manifestly
so has often been noted in connection with Wess-Zumino (WZ) terms associated to
central extensions of a (super)algebra. The WZ term, expressed as an indefinite
integral, is the variable conjugate to the central generator \cite{AL,GGT}. In
our case, we obtain the superpotential term in the action in a
similar way  as the variable conjugate to the $U(1)$ charge
$B$, \emph{even though this charge is not central}. The fact that the
superpotential term in the superspace action \cite{Fub}
cannot be written in manifestly invariant superspace form
leads to a modification of the algebra of Noether charges. This is
in close analogy to the modification of the supertranslation currents for the
super $p$-branes as a consequence of the non-manifest supersymmetry of the WZ
terms in their actions \cite{AIT2}. The analogy is not complete, however, because
in the case under study here the modification can be removed by a redefinition
of the $U(1)$ charge. It is this redefinition that leads to the $\ell$-dependent
expression $B=\hat B + 2\ell$ for the $U(1)$ charge that we mentioned
previously. The mathematical explanation for this $\ell$-dependence is
therefore the non-manifest nature of the superconformal invariance of the
superpotential term in the action. Its physical significance is
best seen in the context of an embedding of the $SU(1,1|1)$ model into the
$SU(1,1|2)$ superconformal model, because of the interpretation of the
(superspace) field equation of the latter as the radial equation for 
a superparticle near an extreme RN black hole.

Many technical aspects of the discussion to follow of the $N=2$ and $N=4$
superconformal mechanics models are similar to 
those in \cite{IKL}, which we became aware of after submission 
to the archives of an earlier version of this paper.
However, the thrust of our argument is quite different, centering as it
does on our improved understanding of the nature and significance of the
superpotential term and the black hole interpretation.    

\section{$N=2$ superconformal mechanics}

We now turn to the $OSp(2|2)$-invariant $N=2$ superconformal mechanics of
\cite{akulov,Fub}. The anticommutation relations of the Lie superalgebra
$osp(2|2)$ are those of (\ref{1.4}). We select ($H$,$Q_i$) ($i=1,2$) as the
`unbroken' generators associated with the real superworldline coordinates
$\zeta^M=(t,\eta^i)$. As before, it is convenient to define $d\tau \equiv
(dt - {i\over 2}d\eta^i \eta_i)$ because we
then have
\begin{equation}
d= d\tau \partial_t + \eta^iD_i\quad ,
\end{equation}
where $D_i$ are the supercovariant derivatives of (\ref{1.3}).

We may write the $OSp(2|2)$ group element as
\begin{equation}
g(t,\eta)= e^{-itH} e^{i\eta^i Q_i} e^{i\lambda^i(t,\eta) S_i}
e^{i z(t,\eta) D} e^{i \omega(t,\eta) K} e^{ i a(t,\eta) B}\, .
\label{3.1}
\end{equation}
Defining $\displaystyle d \xi^M \equiv (d\tau, d\eta_i)$
and $H_A=(H_0,H_i)\equiv (H,Q_i)$ we can rewrite this as
\begin{equation}
i g^{-1} d g = d\xi ^M E_M{}^{~A} \left [ H_A - ({\cal D}_A z) D -
({\cal D}_A \omega) K - ({\cal D}_A \lambda_i) S^i -
({\cal D}_A a) B \right]
\quad,
\label{3.2}
\end{equation}
where $E_M{}^A$ is the worldline supervielbein and ${\cal D}_A$ is the
group-covariant derivative. A calculation yields
\begin{equation}
E_M{}^{~A} = \pmatrix {  e^{-z} & -e^{-z/2} \lambda^T R(a) \cr
                        0      &       - e^{-z/2} R(a)}  \quad ,
\label{3.3}
\end{equation}
where $\lambda^T$ means the transpose of $\lambda^i$ as a two--vector and
$R(a)$
is the $2\times 2$ rotation matrix
\begin{equation}\label{3.4}
R(a) = \pmatrix{ \cos a & -\sin a \cr
                 \sin a & \cos a}
\quad.
\end{equation}
Note that
\begin{equation}
\rm{sdet}\,E = -1 \quad,
\label{3.5}
\end{equation}
so that manifestly invariant actions have the form
\begin{equation}
I = \int dt\,d^2\eta\, {\cal L}({\cal D}_A\phi)
\quad,
\label{3.6}
\end{equation}
where $\phi=(z,\omega,a)$ denotes the set of worldline superfields.
The covariant derivatives can be written as
\begin{equation}
{\cal D}_A = E_A{}^{M} {\cal D}_M\quad,
\label{3.7}
\end{equation}
where $E_A{}^{M}$ is the inverse supervielbein
\begin{equation}
E_A{}^M = \pmatrix { e^{z} & -e^{z} \lambda^T \cr
           0      & -e^{z/2} R^T(a)}
\label{3.8}
\end{equation}
and ${\cal D}_M = ({\cal D}_\tau,{\cal D}_{\eta^i})$ are the
components of the covariant derivatives on the (still non-coordinate) basis
$(d\tau,d\eta^i)$.
The transformation properties of ${\cal D}_M \phi$ are not as simple as those 
of ${\cal D}_A \phi$ (which are superworldline scalars) but they have a 
simpler form. The expressions of ${\cal D}_M \phi$
are found to be
\begin{equation}
\begin{array}{l}
{\cal D}_M z= (2\omega e^{-z} + \dot z\, ,\quad
D_i z -i\lambda_i )
\\[0.3cm]
\displaystyle
{\cal D}_M \omega = (\dot \omega -\omega\dot z - e^{-z} \omega^2 -
{i\over 2} e^z \dot \lambda_i\lambda^i \, ,\quad
D_i\omega -\omega D_i z + i\omega \lambda_i -
{i\over 2} e^z D_i \lambda_j \lambda^j )
\\[0.3cm]
\displaystyle
{\cal D}_M \lambda_j = (e^{z/2} \dot \lambda_k R^k{}_j -e^{-z/2}\omega
\lambda_k R^k{}_j \, ,\quad
e^{z/2} D_i \lambda_k R^k{}_j -e^{-z/2}\omega R_{ij}
+ {i\over 2} e^{z/2} \lambda_i \lambda_k R^k{}_j )
\\[0.3cm]
\displaystyle
{\cal D}_M a = (\dot a -{i\over 2} \lambda_1\lambda_2
\, ,\quad  D_i a -{i\over 2} \epsilon_{ij} \lambda^j ) \quad.
\end{array}
\label{3.8a}
\end{equation}

To proceed, we begin by imposing the manifestly invariant constraint
\begin{equation}
{\cal D}_A z=0 \quad,
\label{3.9}
\end{equation}
which is equivalent to ${\cal D}_M z =0$ and is
solved, algebraically, by
\begin{equation}
\omega= -\frac 12 e^{z} \dot z
\quad,\quad
\lambda_i = -i D_i z
\quad.
\label{3.10}
\end{equation}
As in the previous cases, we could arrange for these equalities to arise as
equations of motion for $\omega$ and $\lambda_i$, but in this case it is
simpler
to impose (\ref{3.9}) as a constraint. As a direct consequence  of
(\ref{3.10}),
we then find that the $A=i$ components of ${\cal D}_A\lambda_j$ satisfy
\begin{equation}
{\cal D}_{(i}\lambda_{j)} =0\ ,
\label{3.11}
\end{equation}
so the \emph{manifestly} superconformal invariant, and $SO(2)$ invariant,
Lagrangian of lowest dimension must be a linear combination of ${\cal D}_0 a$
and $\varepsilon^{ij}{\cal D}_i\lambda_j$. If we insist that our action
describe
the dynamics of a particle in a \emph{one--dimensional} space, with (real)
coordinate $z(t)=z(t,\eta_i)|$, then we cannot make use of ${\cal D}_0 a$. In
this case,
and using that
${\cal D}_A\lambda_j$ for $A=i$ is given by
$\displaystyle e^{z} D_i \lambda_j - \omega \delta_{ij}
+ {i\over 2} e^{z} \lambda_i \lambda_j$,
we get
\begin{equation}
{\cal L} \propto \varepsilon^{ij}{\cal D}_i\lambda_j = -2e^z\left(D_1\lambda_2
+
{i\over 2}\lambda_1\lambda_2\right)\, .
\label{3.12}
\end{equation}
Then, using the constraint (\ref{3.10}), adjusting the proportionality
constant, and integrating by parts, we arrive at the action
\begin{equation}
I_{kin} = {m\over 2} \int dt\, d^2\eta\, e^z D_1 z D_2 z\quad ,
\label{3.14}
\end{equation}
which is the first part of (\ref{1.2}) with $z =\log x^2$.
Let the components of the $z(t,\eta_i)$ superfield be defined by
$z(t)=z|$, $\lambda_i = -i D_i z |$ and $F' = -i D_1 D_2 z |$.
Then, defining new variables $x,\chi,F$ by
\begin{equation}
x=e^{z/2}
\quad,\quad
\chi_i = - \sqrt{{m\over2}} e^{z/2} \lambda_i
\quad,\quad F= 2F'\, ,
\label{3.15}
\end{equation}
and performing the $\eta^i$--integrals, we arrive at a component
action with Lagrangian
\begin{equation}
{\cal L}_{kin} = \left[{m\over 2} \dot x^2 +
{i\over2} (\chi_1\dot \chi_1 + \chi_2\dot \chi_2 ) + {1\over8} m x^2 F^2
+{i\over2} F \chi_1\chi_2\right]\, .
\label{3.16}
\end{equation}

After elimination of $F$ by its algebraic equation of motion the bosonic
Lagrangian reduces to that of (\ref{1.1}) so we have now constructed an
$OSp(2|2)$-invariant extension of the $g=0$ conformal mechanics.  All other
\emph{manifestly} invariant actions must involve either higher-derivatives,
higher powers of first derivatives or (non--auxiliary) bosonic variables other
than $x(t)$. Thus, any $OSp(2|2)$--invariant generalisation of the $g\ne0$
conformal mechanics \emph{cannot be described by a manifestly invariant
action}.
This does not exclude the possibility of an action that is invariant but not
manifestly invariant. The existence of such `non--manifest' invariants has
usually  been associated with the possibility of a central extension of the
Lie (super)algebra of the symmetry (super)group. In such cases the action is
a WZ term (see, for example, \cite{book}). A number of
superworldline examples of this were discussed in \cite{GGT}. In our case,
however, there can be no central extension  because the relevant cohomology of
the $osp(2|2)$ algebra is trivial.
One might therefore be tempted to conclude that there can be no further
$OSp(2|2)$--invariants and hence  that there is no $OSp(2|2)$--invariant
extension of the $g/x^2$ potential of conformal mechanics. But this would be
wrong, as we now explain.

A further $OSp(2|2)$--invariant may be found by the method of \cite{GGT}.
We first note that the bosonic and spinor components of ${\cal D}_A a=({\cal
D}_0 a,{\cal D}_i a)$ are independent superworldline scalar fields because
invariance of the relation $\{D_i,D_j\} = -i\delta_{ij} \partial_t$ requires
the structure group of the frame bundle to be just $SO(2)$.
The group covariant derivatives $\mathcal{D}_A$ transform as a
$SO(2)$ doublet for $A=i=1,2$, so the manifestly $OSp(2|2)$ invariant
constraint
${\cal D}_i a=0$ is also
$SO(2)$ invariant, and equivalent to
\begin{equation}
D_i a - {i\over 2} \epsilon_{ij} \lambda ^j = 0\quad .
\label{3.18}
\end{equation}
Since $\lambda_i = -i D_i z$ (eqn. (\ref{3.10})), this new
constraint implies that
\begin{equation}
D_1 a = \frac 12 D_2 z \quad,
\label{3.19}
\end{equation}
and hence that
\begin{equation}
\dot a = i D_1 D_2 z \quad,
\label{3.20}
\end{equation}
This can be integrated to give
\begin{equation}
a(t) = a| = i \int^t dt'\, D_1D_2 z(t',\eta) |
= i\int^t dt'd^2\eta \, z(t',\eta) \quad .
\label{3.21}
\end{equation}
The variable field $a(t)=a|$  can thus  be viewed as a superspace action in the
form of an indefinite integral. This new action is superconformal invariant, up
to a surface term, because (\ref{3.20}) implies that
\begin{equation}
\delta \left( i D_1 D_2 z \right) = \partial_t {\delta a} \quad .
\label{3.23}
\end{equation}
The left hand side is the variation of the component Lagrangian of the new
superspace action whereas the right hand side is a total time derivative.
By itself, this is not quite sufficient to establish the desired result.
According to (\ref{3.20}), the component Lagrangian is itself a total
time derivative so it is hardly surprising that the same is true
of its variation. Of course, (\ref{3.20}) tells us nothing about the
component Lagrangian; instead it provides us with information about the
independent superfield $a$. However, while $a$ is an independent
superfield its variation $\delta a$ is not. In fact $\delta a$ is a function
of the superfield $z$ and its derivatives, and is independent of
$a$. Thus, we indeed learn from (\ref{3.23}) that the variation of the
component
Lagrangian of the superspace action (\ref{3.21}) is a total time derivative,
and hence that this action is invariant up to a surface term. We have now
deduced that the action
\begin{equation}
I = I_{kin} - 2i\ell \int dt \, d^2\eta\, z\quad ,
\label{3.24}
\end{equation}
where $I_{kin}$ is given in (\ref{3.14}), is $OSp(2|2)$ invariant for arbitrary
real constant $\ell$. Setting $z=\log x^2$ this action is seen to be precisely
that of (\ref{1.2}). The superpotential term is not \emph{manifestly}
invariant because $z$ does not transform as a scalar density. A calculation
shows that $z$ fails to transform as a scalar density by a term that, being
\emph{linear} in $\eta$, does not contribute to
the variation of the superspace integral. Despite the non-manifest
superconformal invariance of the action (\ref{3.24}) the $z$ superfield
equation
\emph{can} be expressed in the manifestly superconformal invariant form
\begin{equation}
\varepsilon^{ij}{\cal D}_i \lambda_j = 4\ell/m
\label{3.24a}
\end{equation}
(recall that ${\cal D}_i$ are the $A=i$ components of ${\cal D}_A$).

The component Lagrangian including the contribution of the superpotential term
(obtained by performing the superspace integrals) is
\begin{equation}
{\cal L} = {1\over 2}m\dot x^2 +
{i\over2} (\chi_1\dot \chi_1 + \chi_2\dot \chi_2 ) + {1\over8} m x^2 F^2
+ F(\ell + {i\over2} \chi_1\chi_2)
\quad.
\label{3.25}
\end{equation}
Elimination of $F$ now yields
\begin{equation}\label{complag}
{\cal L} =
{1\over 2} m\dot x^2 + {i\over 2} (\chi_1\dot\chi_1 +\chi_2\dot\chi_2) -
{2\over mx^2} \ell(\ell + i \chi_1\chi_2)
\quad.
\label{3.26}
\end{equation}
Setting the fermions to zero we recover the bosonic Lagrangian of (\ref{1.1})
with
\begin{equation}
g= 4\ell^2/m \quad.
\label{geee}
\end{equation}
Thus, we have found an $OSp(2|2)$ invariant extension of conformal
mechanics. It is just the model constructed in \cite{akulov,Fub}.
There is an apparent discrepancy with (\ref{ccrel}) but this will be resolved
after we have looked at the $SU(1,1|2)$ model.

\section{$N=4$ superconformal mechanics}

We turn now to the $SU(1,1|2)$-invariant $N=4$ superconformal mechanics
describing the full superparticle radial dynamics. The $su(1,1|2)$ superalgebra
is spanned by the $Sl(2;\bR)$ generators
$(H,K,D)$, the $SU(2)$ generators $J_a$ ($a=1,2,3$), and the
$SU(2)$ doublet supersymmetry charges $(Q^i,S^i)$ and the hermitian
conjugates $(\bar Q_i, \bar S_i)$. The superalgebra has the following
non-vanishing (anti)commutation relations
\begin{equation}
\begin{array}{l@{\qquad}l}
[H,D]=iH \quad, & [K,D]=-iK \quad,
\\[0.3cm]
[H,K]=2iD \quad, & [J_a,J_b] = i\varepsilon_{abc}J_c \quad,
\\[0.3cm]
\{Q^i,\bar Q_j\}= 2\delta^i_j H \quad,
& \{S^i,\bar S_j\}=2\delta^i_j K  \quad,
\\[0.3cm]
\displaystyle \{Q^i,\bar S_j\}= 2(\sigma_a)_j{}^i J_a  + 2i\delta^i_j D
\quad, &
\displaystyle \{\bar Q_i,S^j\}= 2(\sigma_a)_i{}^j J_a  - 2i\delta^i_j D
\quad,
\\[0.3cm]
\displaystyle [D,Q^i]= -\frac i2 Q^i \quad,&
\displaystyle [D,\bar Q_i]= -\frac i2 \bar Q_i \quad,
\\[0.3cm]
\displaystyle [D,S^i]= \frac i2 S^i \quad, &
\displaystyle [D,\bar S_i]= \frac i2 \bar S_i \quad,
\\[0.3cm]
[K,Q^i]= S^i \quad, & [K,\bar Q_i]= -\bar S_i \quad,
\\[0.3cm]
[H,S^i]= Q^i \quad, & [H,\bar S_i]= -\bar Q_i \quad,
\\[0.3cm]
\displaystyle
[J_a,Q^i]= -{1\over2}Q^j(\sigma_a)_j{}^i \quad, &
\displaystyle
[J_a,\bar Q_j]= {1\over2}(\sigma_a)_j{}^k\bar Q_k \quad,
\\[0.3cm]
\displaystyle
[J_a,S^i]= -{1\over2}S^j(\sigma_a)_j{}^i \quad, &
\displaystyle
[J_a,\bar S_j]= {1\over2}(\sigma_a)_j{}^k\bar S_k \quad.
\end{array}
\end{equation}
We take the superworldline-valued supergroup element to be
\begin{equation}
g(t,\eta,\bar\eta) = e^{-itH} e^{i(\eta_iQ^i +\bar\eta^i \bar Q_i)}
e^{i(\lambda_iS^i +\bar\lambda^i \bar S_i)} e^{izD} e^{i\omega K}
e^{i\phi J_1} e^{i\theta J_2} e^{i\psi J_3}
\quad,
\label{eq57}
\end{equation}
where
$\lambda, \bar\lambda,\ z,\ \omega,\ \phi,\theta,\ \psi$
depend on $(t,\eta,\bar\eta)$. The anticommuting coordinates $\eta^i$ and
$\bar\eta_i$ are related by complex conjugation, \emph{i.e.}
$(\eta^i)^*=\bar\eta_i$.
We shall again define
\begin{equation}
d\tau= dt-i(\eta_i d\bar\eta^i+\bar\eta^i d\eta_i) \quad,
\end{equation}
which leads to
\begin{equation}
d=d\tau\partial_t+d\eta_i D^i+d\bar\eta^i\bar D_i\quad,
\label{dconi}
\end{equation}
where
\begin{equation}
D_\eta^i \equiv D^i =
\frac{\partial}{\partial\eta_i}-i\bar\eta^i\frac{\partial}{\partial t}
\quad , \quad
\bar D_{\eta\,i} \equiv
\bar D_i = \frac{\partial}{\partial\bar\eta^i}-
i\eta_i\frac{\partial}{\partial t}
\label{3.61}
\end{equation}
are the superspace covariant derivatives satisfying $\{D^i,\bar D_j\}=
-2i\delta^i_{j} \partial_t$. It should also be noted that $\bar D_i =
-(D^i)^*$.

The left--invariant 1-form can be written as
\begin{eqnarray}
ig^{-1}dg &=& d\xi^M \big[ E_M{}^A H_A - ({\cal D}_M z) D -
({\cal D}_M \omega) K - ({\cal D}_M \lambda)^i S_i
- ({\cal D}_M \bar\lambda)_i \bar S^i \nn
&& - ({\cal D}_M \phi)J_1 -
({\cal D}_M\theta)J_2 - ({\cal D}_M\psi)J_3\big] \quad,
\label{eq58}
\end{eqnarray}
where
\begin{equation}
d\xi^M = (d\tau ,d\eta^i,d\bar\eta_i)\, , \quad
H_A =(H,Q_i, \bar Q^i) \, .
\end{equation}
A calculation yields
\begin{equation}
E_M{}^A =\pmatrix{
e^{-z} & ie^{-z/2}\lambda_l s_j{}^l & -ie^{-z/2}\bar\lambda^l(s^{-1})_l{}^j\cr
0 & -e^{-z/2}s_j{}^i & 0 \cr
0 & 0 & -e^{-z/2}(s^{-1})_i{}^j\cr}\, .
\end{equation}
We shall need the inverse supervielbein
\begin{equation}
E_A{}^M = \pmatrix{e^z & ie^z\lambda_j & -ie^z\bar\lambda^j\cr
0 & -e^{z/2}(s^{-1})_j{}^i & 0\cr
0 & 0 &  -e^{z/2}s_i{}^j\cr}\, .
\end{equation}

Further calculation yields
\begin{equation}
\begin{array}{l}
{\cal D}_M z= \Bigl(2\omega e^{-z} + \dot z\, ,\quad
D^i z +2\bar\lambda^i\, ,\quad c.c.\ \Bigr)
\\[0.3cm]
\displaystyle
{\cal D}_M \omega = \Bigl(
\dot \omega -\omega\dot z - e^{-z} \omega^2 +i(\lambda
\dot{\bar\lambda}+\bar\lambda\dot\lambda)e^z-\frac{1}{36}(\bar\lambda
\sigma\lambda)^2e^z \, ,
\\[0.3cm]
\displaystyle
\quad D^i\omega -\omega D^i z -2\omega \bar\lambda^i -
i(\lambda D^i\bar\lambda+\bar\lambda D^i\lambda) e^z
-\frac{i}{3}e^z(\bar\lambda
\sigma_a\lambda)(\sigma_a\bar\lambda)^i\, ,\quad c.c. \ \Bigr)
\\[0.3cm]
\displaystyle
{\cal D}_M \lambda_i = \Bigl(
(\dot\lambda_k e^{z/2} - \omega e^{-z/2}\lambda_k
 +\frac{i}{6}e^{z/2}(\bar\lambda\sigma_a\lambda)(\lambda\sigma_a)_k)
s_i{}^k \, ,
\\[0.3cm]
\displaystyle
\quad( D^j\lambda_k e^{z/2} - i\omega e^{-z/2}\delta^j_k
-\frac{1}{2}e^{z/2}(\bar\lambda\sigma_a)^j(\lambda\sigma_a)_k-
\frac{1}{2}e^{z/2}\bar\lambda^j\lambda_k) s_i{}^k\, ,
\\[0.3cm]
\displaystyle
\quad(\bar D_j\lambda_k e^{z/2}
-\frac{1}{2}e^{z/2}(\lambda\sigma_a)_j(\lambda\sigma_a)_k+\frac{1}{2}e^{z/2}
\lambda_j\lambda_k)s_i{}^k\ \Bigr)
\end{array}
\end{equation}
and
\begin{equation}
\begin{array}{l}
\\[0.3cm]
\displaystyle
{\cal D}_M \phi=
\Bigl(\dot\phi\cos\theta\cos\psi-\dot\theta\sin\psi+i[Ad(s^{-1})]_{a1}
\bar\lambda\sigma_a\lambda  \, ,
\\[0.3cm]
\displaystyle
 \quad D^i\phi\cos\theta\cos\psi-D^i\theta\sin\psi-2i[Ad(s^{-1})]_{a1}
(\bar\lambda\sigma_a)^i\, ,\quad c.c \ \Bigr)
\\[0.3cm]
\displaystyle
{\cal D}_M \theta= \Bigl(\dot\theta\cos\psi+\dot\phi\cos\theta\sin\psi
+i[Ad(s^{-1})]_{a2}\bar\lambda\sigma_a\lambda  \, ,
\\[0.3cm]
\displaystyle
 \quad D^i\theta\cos\psi+D^i\phi\cos\theta\sin\psi-2i[Ad(s^{-1})]_{a2}
(\bar\lambda\sigma_a)^i\, ,\quad c.c \ \Bigr)
\\[0.3cm]
\displaystyle
{\cal D}_M \psi= \Bigl(\dot\psi-\dot\phi\sin\theta
+i[Ad(s^{-1})]_{a3}\bar\lambda\sigma_a\lambda  \, ,
\\[0.3cm]
\displaystyle
 \quad D^i\psi-D^i\phi\sin\theta-2i[Ad(s^{-1})]_{a3}
(\bar\lambda\sigma_a)^i\, ,\quad c.c \ \Bigr) \quad,
\end{array}
\end{equation}
where the explicit forms of $s_i{}^k$ and $[Ad(s^{-1})]$ are
\begin{equation}
\begin{array}{@{}c}
\displaystyle
s_i{}^k = \left( e^{\frac{i}{2}\psi\sigma_3}
e^{\frac{i}{2}\theta\sigma_2}e^{\frac{i}{2}\phi\sigma_1}\right)_i{}^k
\quad,
\\[0.3cm]
\displaystyle
[Ad(s^{-1})]=\pmatrix
{\cos\theta\cos\psi & -\cos\phi\sin\psi+\sin\phi\sin\theta\cos\psi
 & \sin\phi\sin\psi+\cos\phi\sin\theta\cos\psi\cr
\cos\theta\sin\psi & \cos\phi\cos\psi+\sin\phi\sin\theta\sin\psi &
-\sin\phi\cos\psi+\cos\phi\sin\theta\sin\psi\cr
-\sin\theta & \sin\phi\cos\theta & \cos\phi\cos\theta}
\quad.
\end{array}
\end{equation}

The supercovariant derivatives ${\cal D}_A z$, etc., can now be found from the
formula ${\cal D}_A = E_A{}^M{\cal D}_M$. We begin, as before, by imposing the
manifestly invariant constraint ${\cal D}_Az=0$, which yields
\begin{equation}
\omega = -\frac{1}{2}e^z\dot z\quad ,\quad
\lambda_i ={1\over2}\bar D_i z\, .
\label{2.4}
\end{equation}
This leaves $z$ as the only independent superfield. Manifest $SU(1,1|2)$
invariants will be expressed as full superspace integrals of the form $\int
dtd^4\eta\, {\rm sdet}E{\cal L}$ where ${\cal L}$ is a superworldline scalar,
but
there is now no Lagrangian built from covariant derivatives of $z$ that has a
dimension low enough to yield a kinetic term containing an $\dot x^2$ term.
This
problem could be circumvented by imposing the complex constraint
\begin{equation}
\varepsilon_{ij}{\cal D}^i\bar\lambda^j =0\quad .
\label{compconst}
\end{equation}
The linearisation of this constraint yields $\varepsilon_{ij}D^iD^jz =0$. This
is the reduction to $D=1$ of the $D=4$ `linear' superfield constraint, which is
solved in terms of a conserved vector. The reduction to $D=1$ of a conserved
vector is a triplet $X^i{}_j$ ($X^i{}_i=0$) and a singlet $X$ subject to the
constraint $\dot X=0$. The latter constraint means that the equation of motion
for $X$ (obtained by variation of an action in which $X$ is treated as an
unconstrained superfield) is actually the time-derivative of the true equation
of motion. Thus, the true field equation is the once-integrated $X$-equation in
which there is an arbitrary integration constant. This analysis will apply
equally to the full constraints (\ref{2.4}) except that their solution in terms
of $X^i{}_j$ and $X$ will be more involved\footnote{It was shown in \cite{HST}
how such non-linear constraints may be solved, at least in principle.}. We
thus deduce that the remaining equations of the $SU(1,1|2)$ superconformal
mechanics have the form of an $SU(2)$ triplet equation for $z$ and a singlet
equation involving an arbitrary constant. Both must be constructed from the
supercovariant derivatives ${\cal D}_A\lambda$ and complex conjugates in order
to be manifestly $SU(1,1|2)$-invariant equations of the appropriate dimension.
There is only one candidate for the triplet equation:
\begin{equation}
{\cal D}_{(i}\lambda_{j)} =0\quad .
\end{equation}
The singlet equation is
\begin{equation}
{\cal D}^i\lambda_i + \bar{\cal D}_i\bar\lambda^i = 8\ell/m\quad ;
\end{equation}
as anticipated, it involves an arbitrary integration constant. Choosing the
constant as above, one finds that the bosonic field equation is precisely
equivalent to that derived from (\ref{1.1}), again with $g=4\ell^2/m$.

\section{Superparticle/Black hole interpretation}

We claimed in the introduction that the $N=2$ and $N=4$ models of 
superconformal mechanics describe a particular limit of the 
(planar or full) radial 
dynamics of a superparticle near the
horizon of an extreme RN black hole. In order to justify this claim we must
first account for the discrepancy between the formula (\ref{geee}) for $g$ with
the formula (\ref{ccrel}) found from the superparticle. To do so we must take
into account quantum mechanics. The Hamiltonian corresponding to the Lagrangian
(\ref{complag}) is
\begin{equation}
H= {p^2\over 2m} + {4\ell(\ell + \hat B)\over 2mx^2} \quad,
\label{eq74}
\end{equation}
where
\begin{equation}
\hat B = {i\over2}[\chi_1,\chi_2] \quad .
\label{eq75}
\end{equation}
The phase space Lagrangian is
\begin{equation}
{\cal L}= p\dot x + {i\over2} \delta^{ij} \chi_i\dot\chi_j -H
\quad,
\end{equation}
so that the canonical (anti)commutation relations of the quantum theory are
\begin{equation}
[x,p]= i\, , \qquad \{\chi_i,\chi_j\} = - 2 \delta_{ij}\quad .
\end{equation}
The anticommutation relations are realized by the operators
$\chi_1=i \sigma_1$, $\chi_2 = i \sigma_2$, in which case $\hat B= \sigma_3$.
We see
from this that the eigenvalues of $\hat B$ as an operator in the quantum theory
are $\pm1$. On the $+1$ eigenspace we have
\begin{equation}
H= {p^2\over 2m} + {g\over 2x^2}
\end{equation}
with
\begin{equation}
g= 4\ell(\ell + 1)/m\, ,
\label{eq79}
\end{equation}
which is the $m=q$ case of (\ref{ccrel}). On the $-1$ eigenspace of $B$ we can
take $\ell\rightarrow -\ell$ to arrive at the same result. Thus, our results
are
consistent with those obtained in \cite{claus} once quantum effects are
included
(as they implicitly were in \cite{claus}). We thus confirm the identification
of
the constant $\ell$ in the superconformal mechanics model as the orbital
angular
momentum of a particle near the horizon of a large mass extreme RN black hole.

The operator $\hat B$ is the Noether charge (called $B$ in \cite{Fub}). When
$\ell\ne0$ this is not to be identified with the $U(1)$ charge $B$ in the
superalgebra (\ref{1.4}). Instead, we have
\begin{equation}\label{shift}
B= \hat B + 2\ell \quad .
\end{equation}
That this is a consequence of the non-manifest invariance of the superpotential
term can be seen as follows. The action
\begin{equation}
2 \ell\int dt (\dot a(t) -i D_1 D_2 z| )
\label{shiftb}
\end{equation}
is manifestly invariant, so the Noether charges ${\cal N}$
computed from \emph{this} action by the prescription $\delta I = \int \dot c
\cdot  {\cal N}$  where $c$ is a set of parameters promoted to function of
time,
must close to the algebra of (\ref{1.4}). But $a$ shifts by a constant under
$U(1)$ so that the Noether charge for $I= I_{kin}$ + (\ref{shiftb}) is
now the $B$ of (\ref{shift}). Dropping the $\dot a$ term from
(\ref{shiftb}) leaves us with the actual, but non-manifest, invariant
superconformal mechanics action without the $2\ell$ contribution
to the $U(1)$ charge.
Note that the only additional contribution to the Noether charge from 
$\dot a$ in (\ref{shiftb}) comes from the group variation 
$a'(t')-a(t)$ of the first component of the superfield 
$a(t,\eta_i)$ which is only affected by the $U(1)$ transformations.

As we have shown, the $OSp(2|2)$ invariant superconformal mechanics
is a truncation of an $SU(1,1|2)$ invariant model.
The same is true of the superalgebras; if one sets $Q^2=0$ and
$Q^1=Q$, and 
similarly for $S^i$, and also $J_1=J_2=0$ then one arrives at the algebra of 
$SU(1,1|1)$ in which $(Q,S)$ is the complex $SU(1,1)$ doublet of supercharges 
and $J_3$ is the $U(1)$ charge.
We can now write $Q=Q^1 + i Q^2$ where $Q^i$ are the real supercharges of the 
isomorphic $osp(2|2)$ superalgebra, and similarly for $S$.
Comparison with the $osp(2|2)$ (anti)commuatiton relations given earlierthen 
leads to the identification $2J_3=B$.
Hence,
\begin{equation}\label{orbspin}
J_3 = \ell + {1\over2}\hat B\quad .
\label{eq77}
\end{equation}
If we restrict the dynamics of the particle described by the $SU(1,1|2)$ model
to motion in an equatorial plane then $J_3$ is the particle's angular momentum.
We see from (\ref{orbspin}) that this angular momentum has an orbital component
$\ell$, arising from the presence of the potential term in the action, and a
spin component, arising from the fermion variables.

\vskip 0.5cm

\noindent
{\bf Acknowledgements}:
This research has been partially supported by a research grant PB96-0756
from the DGICYT,  Spain.
PKT thanks for its hospitality the Departamento de
F\'{\i}sica Te{\'o}rica of the University of Valencia, where this work was
begun. JCPB  wishes to thank the Spanish MEC and CSIC for an FPI grant.

\end{document}